\documentclass[%
 aip,
 amsmath,amssymb,
 reprint,%
]{revtex4-1}

\usepackage{graphicx}
\usepackage{dcolumn}
\usepackage{bm}

\usepackage[utf8]{inputenc}
\usepackage[T1]{fontenc}
\usepackage{mathptmx}
\usepackage{etoolbox}

\usepackage{amsmath}
\usepackage{color}
\usepackage{booktabs}
\usepackage{multirow}
\usepackage{comment}
\usepackage{bm}

\DeclareMathOperator{\sech}{sech}

\usepackage{calrsfs}
\DeclareMathAlphabet{\pazocal}{OMS}{zplm}{m}{n}

\makeatletter
\def\@email#1#2{%
 \endgroup
 \patchcmd{\titleblock@produce}
  {\frontmatter@RRAPformat}
  {\frontmatter@RRAPformat{\produce@RRAP{*#1\href{mailto:#2}{#2}}}\frontmatter@RRAPformat}
  {}{}
}%
\makeatother
\begin{document}

\preprint{AIP/123-QED}

\title[Spectrally Informed Learning of Fluid Flows]{Spectrally Informed Learning of Fluid Flows}
\author{Benjamin D. Shaffer}
 \email{ben31@seas.upenn.edu.}

 \affiliation{ 
Department of Mechanical Engineering and Applied Mechanics, University of Pennsylvania, Philadelphia, PA 19104,
USA
}%
\author{Jeremy R. Vorenberg}%
 \affiliation{ 
Air Force Research Laboratory, Albuquerque, NM 87117, USA
}%

\author{M. Ani Hsieh$^1$}

\date{\today}

\begin{abstract}
Accurate and efficient fluid flow models are essential for applications relating to many physical phenomena including geophysical, aerodynamic, and biological systems. While these flows may exhibit rich and multiscale dynamics, in many cases underlying low-rank structures exist which describe the bulk of the motion.
These structures tend to be spatially large and temporally slow, and may contain most of the energy in a given flow. The extraction and parsimonious representation of these low-rank dynamics from high-dimensional data is a key challenge.
Inspired by the success of physics-informed machine learning methods, we propose a spectrally-informed approach to extract low-rank models of fluid flows by leveraging known spectral properties in the learning process.
We incorporate this knowledge by imposing regularizations on the learned dynamics, which bias the training process towards learning low-frequency structures with corresponding higher power.
We demonstrate the effectiveness of this method to improve prediction and produce learned models which better match the underlying spectral properties of prototypical fluid flows.
\end{abstract}

\maketitle

\begin{quotation}
Incorporating prior knowledge into machine learning methods can enhance accuracy, robustness, sample efficiency, and interpretability. Spectral statistics, which describe the energy distribution of a system, are often available even when the physics are not fully understood, arising from fundamental properties like dissipation and large-scale energy injection. Our method, the Spectrally-Informed Autoencoder (SIAE), integrates these spectral properties into Koopman autoencoders (KAEs) by emphasizing low-frequency dynamics with higher power. This integration improves long-term prediction accuracy and results in learned models which better reflect the underlying physics of fluid flows in various scenarios.

\end{quotation}

\section{Introduction}
Accurate and efficient models of fluid flows are essential for a number of applications involving prediction and control such as adaptive optic imaging, agile robotic flight, and  ocean process monitoring. Flows of interest may exhibit complex and nonlinear behavior over a range of scales, thereby posing a key challenge for real-time applications where 
representations which evolve linearly and capture the dominant structure are essential. The infinite dimensional Koopman operator presents a framework to globally linearize dynamics; low-dimensional approximations are therefore highly desirable due to the well-developed tools for analysis and control of linear systems and computational tractability \cite{lusch2018deep}. 
The prospect of learning finite-dimensional approximations of the Koopman operator from data has lead to a surge of recent interest \cite{morton2018deep, salam2022online, rozwood2024koopman, yeung2019learning}, particularly in the form of Koopman autoencoders (KAEs), which combine the representational power of neural networks with the interpretability of the linear Koopman operator \cite{lusch2018deep, takeishi2017learning}.
While purely data-driven approaches to learning these representations can be highly accurate,
in many cases machine learning methods suffer from challenges such as a lack of generalizability, sensitivity to noise, and non-physical features \cite{sharma2023review}.
Physics-informed approaches to machine learning seek to address these shortcomings in a broader class of problems by directly incorporating physical constraints or inductive biases into the learning process \cite{karniadakis2021physics}.
The interpretable form of the linear dynamics in the KAE model readily lends itself to physics-informing techniques, for instance the ability to enforce stability constraints \cite{pan2020physics, erichson2019physics}. In this work, we aim to incorporate general spectral properties of fluid flows as an informative prior to bias the learning process towards models which better correspond to meaningful structures and more closely represent the underlying low-rank structure in flows of interest.

Spectral statistics describing fluid flows arise from first principles or empirical measurement and are widely available, even in cases with multiple or unknown driving physical processes, where conventional physics-informed approaches may not be suitable. Persistent spatio-temporal organizing structures often emerge in fluid flows, which describe the bulk of the motion over a range of flow conditions, enabling reduced-order modeling \cite{taira2020modal}. These structures tend to be spatially large and evolve more slowly than the fluid as a whole.
Many flows of interest, such as fully developed turbulence, are described by a decaying energy spectrum where the largest scales contain the bulk of energy and therefore are most important to capture in a reduced-order model.
Often, this is the result of energy injection at large scales, for instance from boundary conditions, exciting motion, or external radiation, and attenuation at small scales by viscous dissipation to heat.

The aim of this work is to develop and demonstrate a spectral-informing method which incorporates this general principle into the model learning process, specifically, guided by three observations. 1) Fluid flows of interest can be represented by a low-rank description of the motion of persistent, energy-containing structures \cite{taira2020modal}. 2) These dominant structures evolve on relatively slower time scales and larger spatial scales than the fluid as a whole \cite{leonard1975energy}. 3) The evolution of these structures is described by a physically meaningful energy spectrum  \cite{vallis2017atmospheric}.
While these assumptions are not universal, they are generally applicable for simplified modeling of systems of interest, for instance in many geophysical flows.

From these domain specific observations, we propose a Spectrally-Informed autoencoder (SIAE) method built on the KAE architecture. To capture the inherent low-rank structure, we use a lower dimensional embedding, or latent space, resulting in a parsimonious model. The spectrally-informing mechanism encourages identifying coordinates which result in lower frequency dynamics with higher relative power. This is achieved through spectral analysis of the linear dynamics, enabled by the interpretable, linear form of the KAE model dynamics. We aim to construct more models which better match the underlying dynamics by learning dynamics which operate along lower frequencies, expected to correspond to more informative features, and better reflect the relative energy spectrum in the flow of interest.

In the following sections we provide further background to motivate the spectrally-informing mechanism, develop the functionality, and verify the utility of this mechanism on prototypical flows. We demonstrate the success of both the spectral-informing and low-frequency bias terms in constructing models which produce more accurate long term predictions, with the additional desirable properties of operating on lower frequencies and closer matching the known spectrum. This approach provides a general and robust method for imparting physical knowledge in the form of known spectral properties directly into the learned dynamics of data-driven models, which could be applied on a range of problems that match our underlying assumptions.

\section{Background}
Numerical solutions to the incompressible Navier-Stokes equations describing fluid motion are computationally expensive.
As a result, faster reduced-order models (ROMs) are essential for many applications in fluid dynamics involving real time sensing or control, such as distributed robotic sensing \cite{salam2020adaptive} and predictive adaptive optics control \cite{shaffer2023generalizable}, among others. 
Advances in data availability and computational capabilities allow for a data-driven approach to constructing ROMs, either via linear methods such as proper orthogonal decomposition (POD) \cite{berkooz1993proper} or more representational autoendoer (AE) neural networks \cite{loiseau2018sparse}. Parsimonious, or sparse, representations may result in more general and physical models \cite{kutz2022parsimony}. 

Fluid applications of machine learning are particularly interesting due to the complex nature of the flow and well studied equations of motion. 
ML-aided approaches to fluid modeling have been applied to address a wide range of challenges such as turbulence closure, flow prediction, and parameter estimation  \cite{vinuesa2022enhancing, kochkov2021machine, pandey2020perspective, duraisamy2019turbulence, ahmed2021closures, brunton2020machine, sharma2023review}. 
A new physics-informed approach \cite{raissi2019physics} has demonstrated substantial potential for solving partial differential equations, particularly by enforcing the momentum and mass conservations described by the Navier Stokes equations in the loss function of a neural network \cite{cai2021physics, sharma2023review}.
Often, the predictions from these models are evaluated against the known statistical properties, such as the Kolmogorov Spectrum for turbulent flows.

There has been broad recent interest in the Koopman operator framework for various applications, particularly involving control of nonlinear systems \cite{kaiser2021data, korda2018linear}. Originating in fluid dynamics, Dynamic Mode Decomposition (DMD) is a method for identifying dominant modes of oscillation from spatiotemporal data \cite{schmid2022dynamic}, which can be viewed as a linear approximation of the Koopman modes \cite{rowley2009spectral}. While this approximation can be useful, it is also sensitive to slight nonlinearities and noise \cite{wu2021challenges}. The Extended DMD (EDMD) framework introduces nonlinear lifting by linear regression over a library of candidate functions \cite{li2017extended}, however this relies on a well chosen library.
The KAE architecture we adopt can be seen as a generalization of DMD with a learned general encoding as lifting. Neural networks have been used to learn approximate representations of the Koopman eigenfunctions for some time \cite{takeishi2017learning}. Our work most closely follows the method developed by Lusch et al. \cite{lusch2018deep} due to the parsimonious representation and interpretability of the low-rank latent space, however we do not use the proposed continuous eigenvalue parameterization network.
Physics-informed approaches to the Koopman autoencoder include enforcing temporally consistent predictions \cite{nayak2024temporally, azencot2020forecasting}, limiting learning to dynamics on an attractor \cite{constante2024data}, and implicit \cite{erichson2019physics} or explicit stability constraints \cite{pan2020physics}. Similar to our approach, Miller et al. utilize regularization on the eigenvalues of the learned linear operator \cite{miller2022eigenvalue, naiman2023generative}. Pan et al. \cite{pan2020physics} propose learning the residual of a DMD model, and suggest using DMD for initialization. Typically, these methods are general developments which address the stability and persistence of the learned Koopman model. In contrast, our novel SIAE method utilizes domain specific insights to inform the frequency content of the KAE model.

While several recent works have addressed the energy spectrum of fluid flow models with PINNs \cite{mi2023cascade}, these approaches do not typically incorporate the spectrum as an inductive bias of the model. Recent work extracts statistics describing the turbulent regime from limited measurement \cite{buzzicotti2022inferring}, which forms a compliment to our work where underlying statistics are used to direct learning.

\section{Methodology}

\subsection{Problem Setup}
We assume the system of interest can be modeled as a discrete time, autonomous, dynamical system in the form
\begin{equation}
    \mathbf{z}_{k+1} = \pazocal{A}(\mathbf{z}_k)
\end{equation}
where $\mathbf{z}_{k}\in\mathbb{R}^n$ indicates the state of the system at time $k\in\mathbb{Z}^+$ and the function $\pazocal{A}: \mathbb{R}^n\rightarrow\mathbb{R}^n$ maps the state of the system forward in time from $k$ to $k+1$. We assume the dynamics in $\pazocal{A}$ evolve $\mathbf{z}$ in a low-dimensional state space governed by the evolution of dominant flow structures \cite{brunton2019data}.
In application, the dynamics in $\pazocal{A}$ are often unknown, further, the low-dimensional state is often observed by projection onto a higher dimensional subspace $\pazocal{X}$, such that,
\begin{equation}
    \label{eq:observation}
    \mathbf{x}_{k} = \pazocal{H}(\mathbf{z}_k) + \eta
\end{equation}
where $\mathbf{x}_{k}\in\mathbb{R}^n$ is the observed state at time $t$ with measurement error $\eta$, and $\pazocal{H}: \mathbb{R}^n\rightarrow\mathbb{R}^m$ is an unknown sampling function that maps from the state space to a snapshot. It is assumed that the dimension of the observations is much larger than the dimension of the underlying system, $n<<m$, and that $\pazocal{H}$ is invertible. The snapshots then evolve according to $\pazocal{F}:\mathbb{R}^m\rightarrow\mathbb{R}^m$ as, $\mathbf{x}_{k+1} = \pazocal{F}(\mathbf{x}_k)$. The problem setup is shown in Figure \ref{fig:problem setup chart 1} (a). In practice, it is assumed we have access to only the high-dimensional snapshots from experimentation, for instance video recording of a flow field, and the low rank state must be recovered from the available measurements.

This formulation encapsulates two of the key challenges in data-driven dynamical systems: experimentation often provides high-dimensional representations of the low-rank state variables, and the underlying equations are unknown.
While the exact representations of the system may not be known, general properties often are, particularly in the form of spectral statistics.
By incorporating this knowledge into model training, we effectively restrict the space of solutions, which may result in more data-efficient training and robust models.

\subsection{Koopman Operator Theory}

Following the recovery of a low-rank state from high-dimensional measurements, we leverage the Koopman operator framework to construct a linear model that captures the time evolution of these states.
The Koopman operator perspective  provides an alternative description of dynamical systems in terms of the evolution of the Hilbert space of possible measurements given by $y = g(\mathbf{z})$ \cite{koopman1931hamiltonian}.
The Koopman operator, denoted $\pazocal{K}$, is an infinite dimensional linear operator that advances these functions of the state in time,
\begin{equation}
    \pazocal{K} g = g \circ \pazocal{A}
\end{equation}
where $\circ$ is the composition operator. The evolution of the states is then described by 
\begin{equation}
    \label{eq: Koopman evolution}
    g(\mathbf{z}_{k+1}) = g(\pazocal{A} (\mathbf{z}_k)) = \pazocal{K} g(\mathbf{z}_k).
\end{equation}

\begin{figure}
    \centering
    \includegraphics[width=0.65\columnwidth]{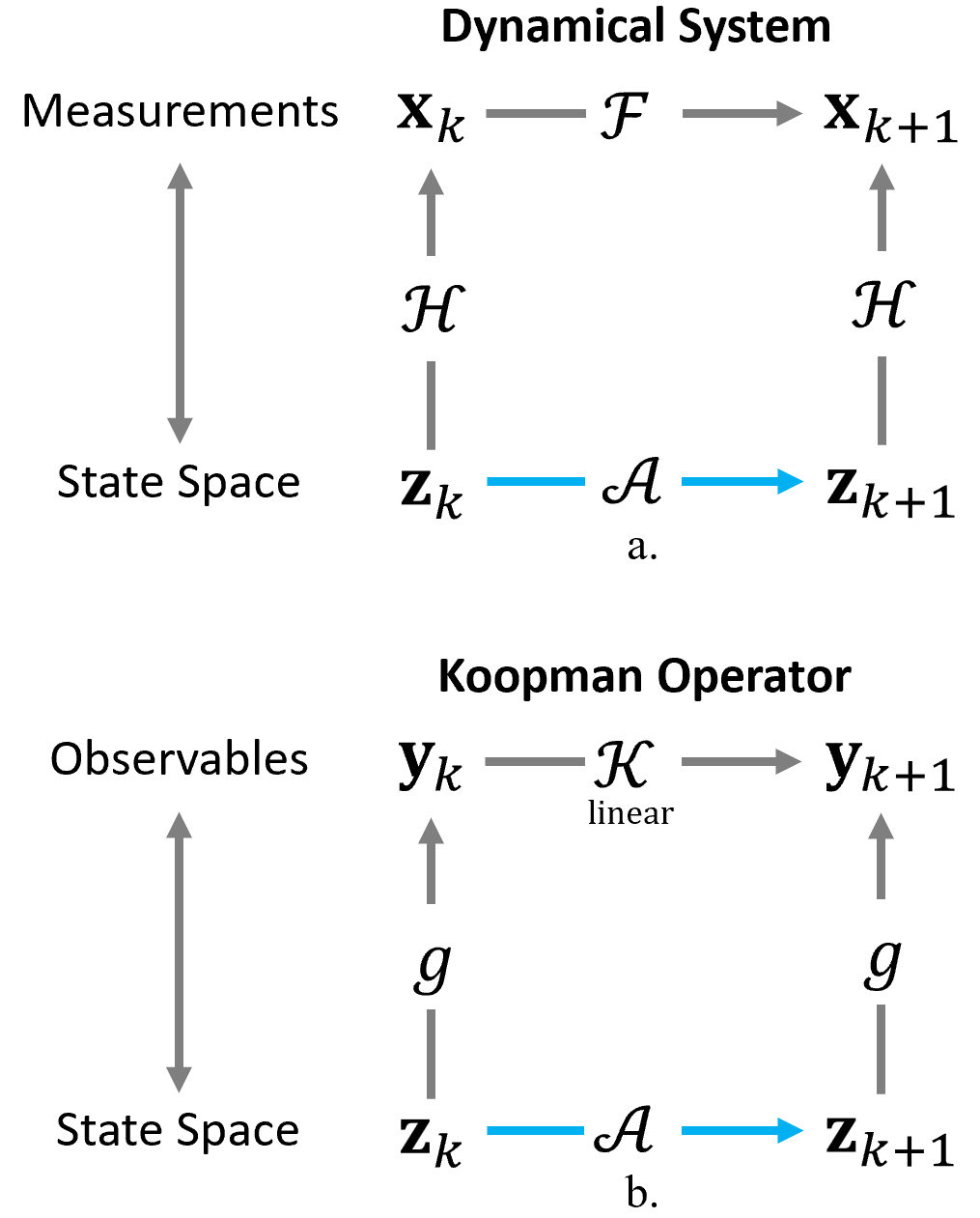}
    \caption{We aim to capture low-rank, linear models describing complex systems by exploiting underlying simplicity. We consider a problem where the low-rank state space is sampled to higher dimensional measurements in (a), which can be modeled by the infinite dimensional linear Koopman operator on the space of observables in (b). The blue arrows emphasis the evolution of the underlying dynamical system of interest. In both subfigures the space with larger dimension is placed above.}
    \label{fig:problem setup chart 1}
\end{figure}

The linearity of the Koopman operator enables use of existing linear system analysis techniques, but the infinite dimension is intractable to represent or compute \cite{brunton2019data}. The eigenfunctions of the Koopman operator provide intrinsic coordinates over which the key measurement functions evolve linearly, allowing for truncated representations. While systems  may permit exact finite dimensional models, applied Koopman theory typically considers an approximation of leading eigenvectors \cite{cohen2024functional}. The eigen-decomposition of the Koopman operator allows for a compact and interpretable representation. For a discrete time Koopman operator with eigenfunction $\varphi(\mathbf{z})$ and associated eigenvalue $\lambda$, the evolution of snapshots is expressed as
\begin{equation}
    \varphi(\mathbf{z}_{k+1}) = \pazocal{K}\varphi(\mathbf{z}_k) = \lambda\varphi(\mathbf{z}_k).
\end{equation}
Crucially, the eigenfunction coordinates are sufficient to globally linearize the strongly non-linear dynamics, and evolve according to the associated eigenvalues \cite{bevanda2021koopman}. A description of the Koopman operator framework is shown in Figure \ref{fig:problem setup chart 1} (b).

\subsection{Learning Encoded Dynamics}
\begin{figure}
    \centering
    \includegraphics[width=0.65\columnwidth]{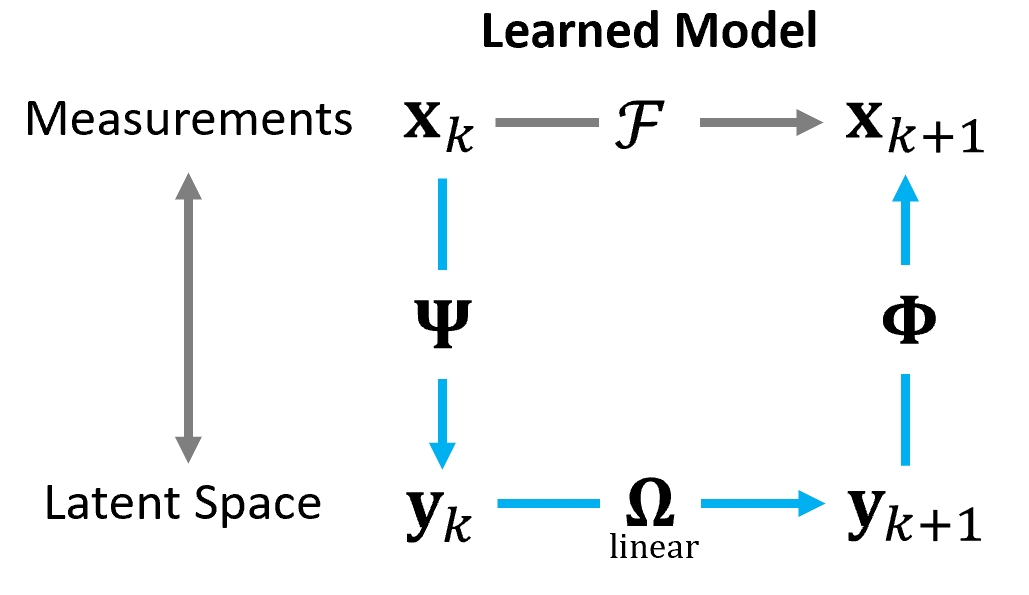}
    \caption{We consider a data-driven approach to extracting  underlying low-rank linear models that capture the evolution of persistent, energy containing structures in high-dimensional flows. This involves approximating the state space in a so-called latent space and approximating the underlying dynamics with a linear map, implicitly learning both transformations in Figure \ref{fig:problem setup chart 1}. In this work we investigate using known spectral properties of the underlying dynamical system to improve the learning of $\bm{\Omega}$.}
    \label{fig:problem setup chart 2}
\end{figure}

The aim of this work is therefore to use a sequence of high-dimensional snapshots,
\begin{equation}
    \mathbf{X} =
        \begin{bmatrix}
            \mathbf{x}_1, & \mathbf{x}_2, & \cdots & \mathbf{x}_T
        \end{bmatrix}
\end{equation}
to extract a linear model approximating the underlying low-rank dynamics $\bm{\Omega}: \mathbb{R}^h \rightarrow \mathbb{R}^h$, which approximates the Koopman operator, $\pazocal{K}$, acting on the state space which evolves according to the nonlinear dynamics in $\pazocal{A}$. 
This entails learning the mapping from snapshots to low-dimensional coordinates spanned by a set of Koopman eigenfunctions, and the inverse of this mapping. The coordinates form the latent space of the neural network with dimension $h$. The encoding function $\bm{\Psi}: \mathbb{R}^m \rightarrow \mathbb{R}^h$ maps from snapshot to latent space, and the decoding function $\bm{\Phi}: \mathbb{R}^h \rightarrow \mathbb{R}^m$ maps back from latent space to snapshot. The state of the latent space at time $k$ is denoted $\mathbf{y}_k$ therefore,
\begin{equation}
    \mathbf{y}_k = \bm{\Psi}(\mathbf{x}_k)
    \quad\text{and}\quad 
    \mathbf{\hat{x}}_k = \bm{\Phi}(\mathbf{y}_k).
\end{equation}
The linear dynamics in $\bm{\Omega}$ evolve the state in the latent space $t$ steps forward in time as
\begin{equation}
    \mathbf{\hat{y}}_{k+t} = \bm{\Omega}^t \mathbf{y}_k,
\end{equation}
we will simplify the notation by assuming $t=1$.
The KAE model is formulated as the composition of these operations, $\pazocal{F}(\cdot) \approx \bm{\Phi} \circ \bm{\Omega} \circ \bm{\Psi}(\cdot)$, such that
\begin{equation}
    \mathbf{\hat{x}}_{k+1} = \bm{\Phi} \circ \bm{\Omega} \circ \bm{\Psi}(\mathbf{x}_k).
\end{equation}
The description learned model is shown schematically in Figure \ref{fig:problem setup chart 2}. The dimension of the state space, $n$, is assumed to be unknown; the dimension of the latent space of the learned model, $h$, must be carefully selected to capture meaningful features while suppressing uninformative variance \cite{erichson2019physics}.

The autoencoder model is learned in a data-driven fashion over $\mathbf{X}$ by minimizing the $\pazocal{L}_2$ norm of the prediction error between a pair of snapshots $\{\mathbf{x}_k, \mathbf{x}_{k+1}\}_{t=1,2,\dots}$, for a given time step
\begin{equation}
    \pazocal{L}_{prediction,\ k}={\lVert \mathbf{x}_{k+1} - \bm{\Phi} \circ \bm{\Omega} \circ \bm{\Psi}(\mathbf{x}_k)\rVert_2^2.}
\end{equation}
To ensure that the dynamics are entirely responsible for the temporal evolution of the state, it must be enforced that
\begin{equation}
    \label{eq: identity of encoder and decoder}
    \mathbf{x}_k \approx \bm{\Phi}\bm{\Psi}(\mathbf{x}_k)
    \quad\text{and}\quad 
    I \approx \bm{\Phi}\bm{\Psi}.
\end{equation}
which can be accomplished with a reconstruction loss term \cite{lusch2018deep} for a given snapshot as,

\begin{equation}
    \label{eq: Reconstruction loss}
    \pazocal{L}_{recon,\ k}={\lVert \mathbf{x}_{k} - \bm{\Phi} \circ \bm{\Psi}(\mathbf{x}_k)\rVert_2^2.}
\end{equation}

\begin{figure*}
    \centering
    \includegraphics[width=1.25\columnwidth]{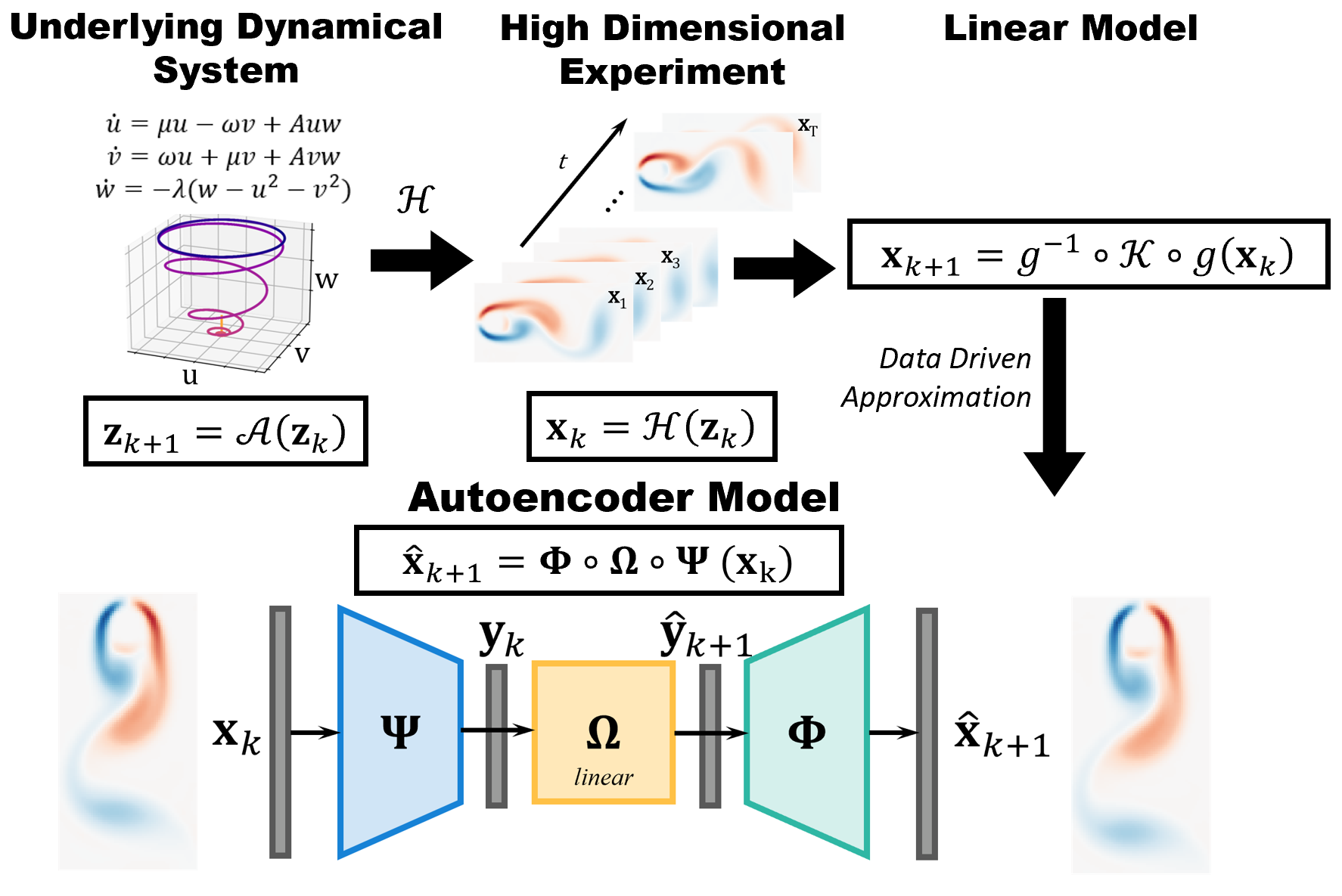}
    \caption{Visual description of problem setup, using periodic vortex shedding as a sample system. While the evolution can be described in only three dimensions, DNS data may have $O(1e6)$ or greater dimensions. Our goal is to extract a low-rank linear model which approximates the underlying dynamics using a KAE structure. Isolating the linear dynamics enables the spectral-informing mechanism described in this work.}
    \label{fig:vizual schematic}
\end{figure*}

Combining the reconstruction and prediction loss terms over the length of the data set gives the base KAE loss function
\begin{equation}
\begin{split}
    \label{eq: AE loss}
    \pazocal{L}_{AE} = \frac{1}{T-1}\sum^{T-1}_{k=1} & \lVert \mathbf{x}_{k+1} - \bm{\Phi} \circ \bm{\Omega} \circ \bm{\Psi}(\mathbf{x}_k)\rVert_2^2 \\
    & + \alpha \lVert \mathbf{x}_{k} - \bm{\Phi} \circ \bm{\Psi}(\mathbf{x}_k)\rVert_2^2 
\end{split}
\end{equation}

where $\alpha$ is a weighting parameter to balance the two objective, in these results we use $\alpha=2$ for all cases. An overview of the model architecture and problem framework is given in Figure \ref{fig:vizual schematic}.

The model architecture is constructed to mirror the problem setup through the discovery of low-rank dynamics from high-dimensional flow snapshots.
We can view the autoencoder learning process described in eq. \eqref{eq: AE loss} as learning a finite dimensional approximation of the Koopman operator using a limited set of measurements described by the functions $\bm{\Phi}$ and $\bm{\Psi}$. Therefore, the eigenvectors and eigenvalues of the linear dynamics in $\bm{\Omega}$ approximate those of $\pazocal{K}$, and spectral regularizations applied on the learned dynamics can be viewed as enforcing intuitions about the expected properties of the Koopman operator.

The autoencoder model is agnostic to the architecture of the encoder and decoder, provided there is no state memory that would corrupt the temporality of the input/output structure. This property arises from the identity enforcing loss in eqs. \eqref{eq: identity of encoder and decoder} and \eqref{eq: Reconstruction loss}, and enables a range of application dependent state-of-the-art implementations. Although it is not examined in this work, the encoder and decoder are also excellent candidates for pretraining on related tasks where ample data is available, particularly due to the apparent universality of certain structures in fluid flows \cite{taira2017modal}.

\section{Spectrally-Informing Learned Dynamics}
The KAE approach yields a linear approximation of the underlying dynamics, enabling a novel spectral-informing approach based on the diagonalization of the linear operator. In this section we describe our implementation, as well as the previously developed stability constraint, all of which rely on the interpretable linear dynamics. Our spectral-informing approach has two objectives: to learn dynamics which operate along lower frequencies, which are expected to correspond to more general and informative features in the flow, and to match the relative power at these frequencies in the model's latent space to a prior known power spectrum. We propose that by combining these should result in a model which best matches the flow of interest in applications matching our assumptions.

The autoencoder formulation approximates the underlying nonlinear dynamics in $\pazocal{A}$ with a time-invariant linear map $\bm{\Omega}$, which can be diagonalized into its eigenvalues and eigenvectors as $\bm{\Omega} = \Phi\Lambda \Phi^{\dagger}$, where $\Phi^{\dagger}$ is the psuedoinverse of $\Phi$.
The mapping forward in time can be expressed as,
\begin{equation}
    \label{eq: eigen expansion of time evolution}
    \mathbf{x}_{k+1} = \sum_{j=1}^h\varphi_j\lambda_j b_{kj}
\end{equation}
where $\mathbf{b}_k = \Phi^\dagger\mathbf{x}_k$ are the amplitudes associated with a given snapshot projected on the eigenvectors by the psuedoinverse. A given eigenvalue can be expressed as $\lambda_j=\exp{s_j}$, such that $s_j = \gamma_j + i\omega_j$, yielding growth/decay $\gamma_j$ and eigenfrequency $\omega_j$ \cite{mezic2013analysis}.
In discrete space, the eigenfrequency equates to the phase and growth/decay corresponds to the magnitude of a given eigenvalue, with a magnitude of 1 indicating persistence \cite{rowley2009spectral}. From the frequencies in $\Omega=[\omega_1,\omega_2,\cdots,\omega_n]$ and amplitudes in $\mathbf{b}$ we can construct a discrete power spectrum, describing the evolution applied by the linear operator on a given dataset. Using the averaged amplitudes normalized to unit energy, computed for a given frequency indexed by $j$ as,
\begin{equation}
    \langle\mathbf{b}\rangle_j = \frac{\sum^p_{k=1}{\lVert b_{kj}\rVert_2}}{\sum^h_{j=1}{\sum^p_{k=1}{\lVert b_{kj}\rVert_2}}}
\end{equation}
where $p$ is the minibatch length, or number of samples processed together in one iteration during training, we express the discrete energy spectrum of the operator as
\begin{equation}
    E_{\bm{\Omega}} =\{(\omega_i, \langle\mathbf{b}\rangle_i)\ | i \leq h, i \in \mathbb{Z}^+ \}.
\end{equation}
During training this spectrum is computed over each minibatch, with a batch size of 32 used in all cases. 

We can inform the operator spectrum by regularization when a \emph{a-priori} physical spectrum describing the evolution of measurements is known, for instance one following the form of the Kolmogorov hypothesis.
We refer to the \emph{a-priori}, or physics-based, spectrum as $E_{prior}(f) = \sigma$.
For comparison, the prior spectrum is normalized to unit energy at the eigenfrequencies as,
\begin{equation}
    \Bar{E}_{prior} = E_{prior}/\sum_{i=1}^h E_{prior}(\omega_i)
\end{equation}
We define a spectral loss term which penalizes a discrepancy between the learned and prior spectra:
\begin{equation}
    \label{eq: spectral loss}
    \pazocal{L}_{spectral} = \frac{1}{h}\sum_{i=1}^h{\lvert \Bar{E}_{prior}(\omega_i) - \langle\mathbf{b}\rangle_i \rvert^2}.
\end{equation}
This loss term is minimized when the discrete operator spectrum has the same relative power at each eigenfrequency as the informing prior spectrum.

\begin{figure}
    \centering
    \includegraphics[width=1.0\columnwidth]{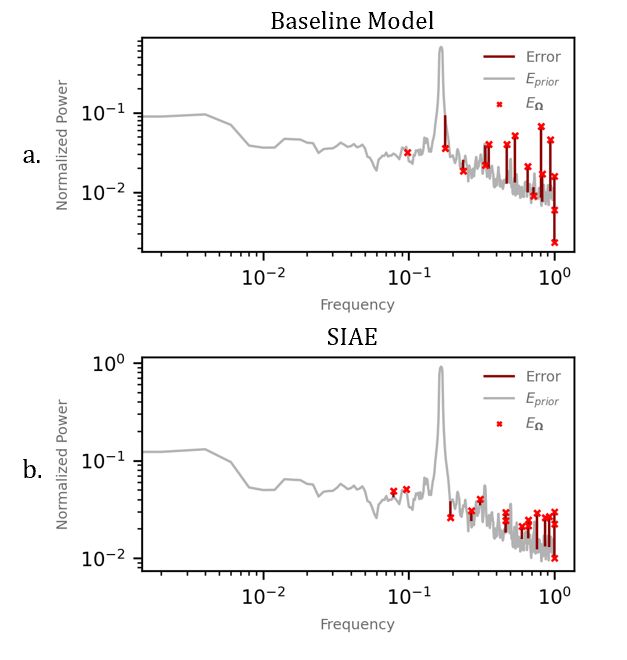}
    \caption{The spectral loss is computed as the average difference between the discrete operator spectrum and a prior spectrum, both normalized for comparison. This graphic shows the final operator spectrum in red \textit{x}'s overlaid on the prior spectrum which was determined by taking the spatially averaged temporal power spectrum of sea surface temperature in the Gulf of Mexico for (a) the baseline model and (b) the model with a spectral regularization term. The spectrally-informed model more closely matches the underlying properties, particularly by reducing the power at higher frequencies. }
    \label{fig:spectral comp}
\end{figure}

The effect of this regularization term is demonstrated on Sea Surface Temperature (SST) data in Figure \ref{fig:spectral comp}. We observe that while the baseline model systematically over-represents the energy at high frequencies in the latent space, the inclusion of the spectral regularization results in a model which more closely matches the energy content of the informed prior, in this case computed from the training snapshots.

The spectral-informing mechanism is flexible to a range of applications based on the choice of the prior. In all the examples in this work, we determine $E_{prior}$ by taking the spatially average temporal power spectrum of the data snapshots over the training samples. This can be viewed as a general feature extraction technique which is enforced in the latent space. In many applications there are known spectra which can be directly enforced by the spectral-informing mechanism, such as the Kolmogorov spectrum for 3D turbulence,
or the JONSWAP spectrum describing the distribution of oceanic waves \cite{hasselmann1973measurements}. Further, this methodology could be applied to achieve specific engineering goals, for instance in the case of control with limited bandwidth, it may be desirable to construct models with a decaying power spectrum at increasing frequencies.

The spectral decomposition in eq. \eqref{eq: eigen expansion of time evolution} provides that the temporal evolution of the operator is described along the eigenvalues and therefore the associated eigenfrequencies.
The spectral-informing mechanism encourages matching relative power to the prior known spectrum at a given frequency, but does not directly influence the learned frequencies. Broadly, features of interest are spatially large and temporally slow. Therefore lower eigenfrequencies are expected to describe more informative and energy containing features in a given flow.
While, it is possible that this is achieved implicitly in the learning process and minimization of the spectral loss in eq. \eqref{eq: spectral loss}, we consider explicitly incentivizing lower eigenfrequencies via an additional loss term, which acts as a regularization on the dynamics in $\bm{\Omega}$.
The low-frequency bias takes the form
\begin{equation}
    \pazocal{L}_\omega = \frac{1}{h}\sum_{i=1}^h{\lvert\omega_i\rvert^2},
\end{equation}
which is minimized when all eigenfrequencies are zero and therefore must be balanced with the data driven training objectives in eq. \eqref{eq: AE loss}.

\begin{figure*}
    \centering
    \includegraphics[width=1.6\columnwidth]{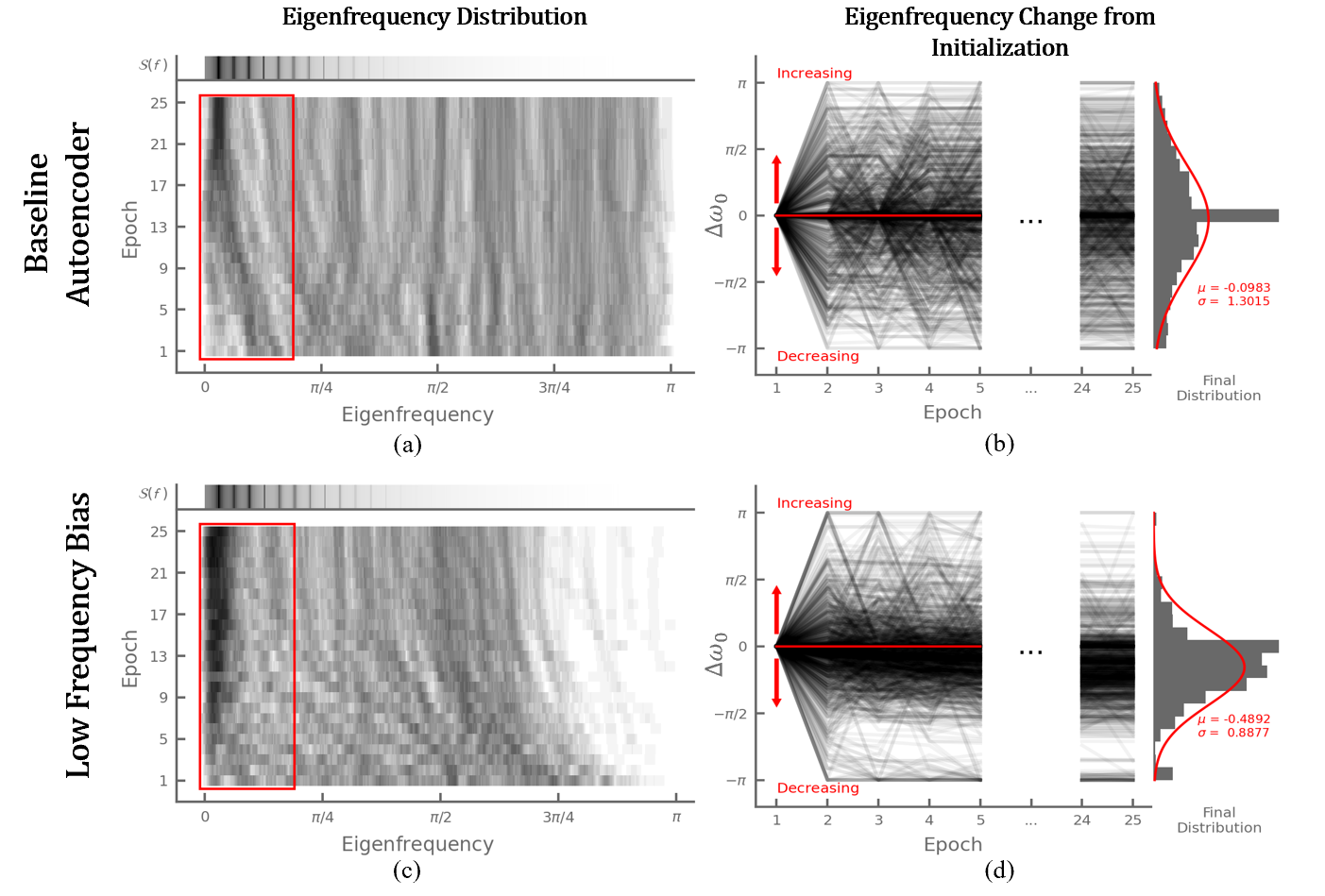}
    \caption{Comparison of eigenfrequency distribution for 64 models over 25 epochs of training on vortex shedding data for (a) baseline model and (b) low-frequency biased model, and changes from initialization of eigenfrequencies for (c) baseline model and (d) low-frequency biased model. The red region in (a) and (c) highlight the consolidation of low eigenfrequencies to the dominant energy peak during training, the data spectrum is given for comparison ($S(f)$). }
    \label{fig:low f bias demo}
\end{figure*}

Figure \ref{fig:low f bias demo} demonstrates the effect of the low-frequency bias on the distribution of eigenfrequencies of the learned linear operator  describing flow past a cylinder during training over 64 trials.
This example is periodic, resulting in peaks in the power spectrum at the natural frequency and harmonics. We observe that while the baseline model does succesfully identify the natural frequency  (indicated with a red region in Figure \ref{fig:low f bias demo}) higher frequencies with low power are overrepresented in the final model. By introducing the low-frequency bias, we observe a systematic shift towards lower frequencies during training, which may better represent the underlying spectrum.

\begin{figure}
    \centering
    \includegraphics[width=1.0\columnwidth]{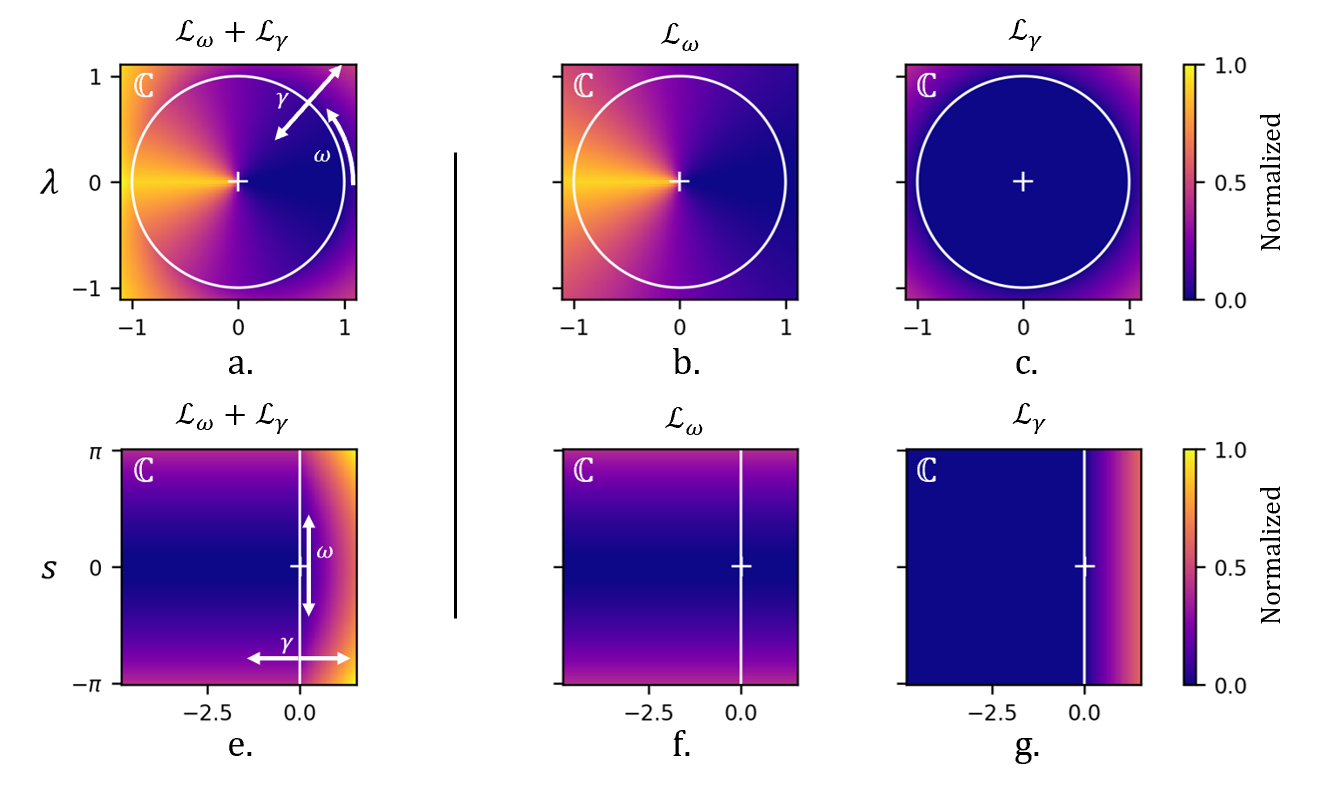}
    \caption{Total loss assigned to eigenvalues in (a) discrete time $\lambda$, and (e) continuous time $s = ln(\lambda)$, with low-frequency loss $\pazocal{L}_\omega$ shown in (b) and (f), and stability loss $\pazocal{L}_\gamma$ shown in (c) and (g). A brighter color corresponds to higher loss; a model with low eigenvalue regularization loss would have all eigenvalues at values closer to zero loss, which correspond to low-frequency and stable dynamics.}
    \label{fig:eig losses}
\end{figure}

Following Erichson et al. \cite{erichson2019physics} we also introduce a term to encourage stable solutions; a linear map is stable in the sense of Lyapunov if all continous time eigenvalues have a negative real component, or $\gamma_i < 0, \forall i \in \{1,2,\dots,h\}$ \cite{sastry2013nonlinear}.
The stability loss term is implemented as
\begin{equation}
    \pazocal{L}_\gamma = \frac{1}{h}\sum_{i=1}^h{\rho(\lvert\gamma_i\rvert)}.
\end{equation}
where
\begin{equation}
    \rho(x) = \begin{cases} 
        x & \text{if } x > 0 \\
        0 & \text{otherwise} 
        \end{cases}
\end{equation}
The low-frequency bias and stability loss terms are independently applied on the real, $\gamma$, and imaginary, $\omega$, axes of the continuous eigenvalues, $s$. We visualize the normalized loss values associated with eigenvalues in the complex plane in Figure \ref{fig:eig losses}, for both continuous and discrete spectra.

To construct our Spectrally-Informed autoencoder (SIAE) model we combine the physical loss functions with the base autoencoder loss described in eq. \eqref{eq: AE loss} as

\begin{equation}
    \label{eq: SIAE loss}
    \pazocal{L}_{SIAE} := \pazocal{L}_{AE} + \alpha_{spectral}\pazocal{L}_{spectral} + \alpha_{\omega}\pazocal{L}_{\omega} + \alpha_{\gamma}\pazocal{L}_{\gamma}
\end{equation}
where $\alpha_{spectral}$, $\alpha_{\omega}$, and $\alpha_{\gamma}$ are tunable parameters balancing the various objectives.
For spectra where the assumptions underlying the development of these methods hold, such as in fully developed turbulence, we expect the spectral-informing mechanism and low-frequency bias to have complimentary effects, i.e. the combination of both terms results in the best model performance.

\section{Results}

\subsection{Data overview}
\begin{figure}
    \centering
    \includegraphics[width=0.9\columnwidth]{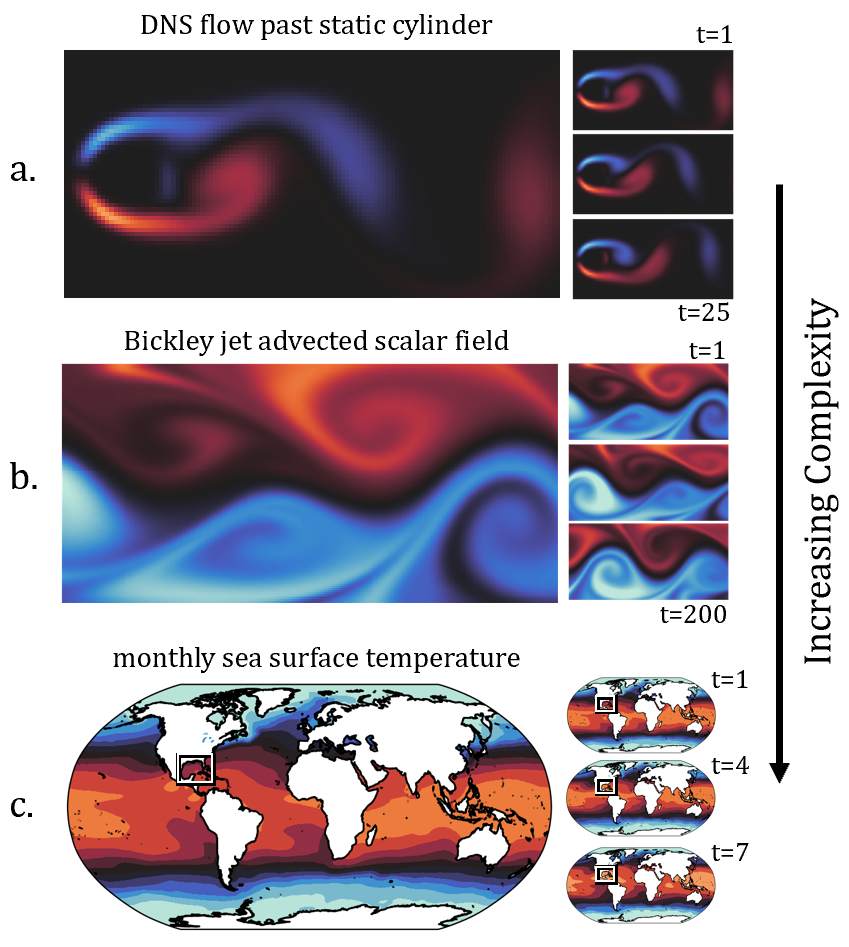}
    \caption{Overview of representative systems analyzed in this study, for each we give a detailed snapshot ($\mathbf{x}_0$) and a sample time evolution to demonstrate how the systems evolve. The three datasets are (a) periodic vortex shedding, (b) meandering zonal jet or Bickley jet, and (c) monthly sea surface temperature in the Gulf of Mexico (windowed region). These cover a range of conditions and increasing complexity, from DNS to experimental and Reynolds number from $Re\approx160$ to Earth scale ($Re>1e6$). In each we consider a scalar field evolving over time.}
    \label{fig:data overview}
\end{figure}

To demonstrate the empirical performance of the spectral-informing approach we consider three prototypical examples of fluid flows with increasing complexity; Direct Numerical Simulation (DNS) of laminar flow past a cylinder, a scalar field advected by Bickley jet flow, and monthly SST measurements in the Gulf of Mexico. Sample snapshots from each dataset are shown in Figure \ref{fig:data overview}. In addition to representing cases of increasing complexity and scale, these examples also cover a range of physical assumption, highlighting the generalizability of this method beyond cases directly modeled by the Navier Stokes equations.

\subsubsection{Periodic vortex shedding}
We consider two-dimensional flow past a cylinder at $Re=160$, which is characterized by periodic, laminar, vortex shedding. This is a canonical benchmark dataset and has been well analyzed in previous work, although it is much simpler than typical flows of interest \cite{callaham2019robust}. 
By truncating instabilities in time, we consider only the stable oscillatory manifold describing the limit cycle of the system.
Additionally, the data was processed by dynamic mode decomposition (DMD) to give an extended and noise-free trajectory; the DMD model is entirely linear and is highly accurate for describing oscillations on the limit cycle. Linear Koopman estimation techniques such as DMD can therefore result in an exact model for this data.

The data were retrieved from a publicly available repository provided by the ETH Zurich Computer Graphics Laboratory \cite{Guenther17}. The 2D simulation was computed using Gerris Flow Solver \cite{gerrisflowsolver} and resampled to a regular grid, full details are available in Guenther et al. \cite{Guenther17}. We compute the vorticity as the curl of the velocity field to give a scalar value, and use a windowed section of the full data nearest to the cylinder. The model trained on 500 snapshots and evaluated on 500 separate snapshots. Model predictions were evaluated over a horizon of 100 future time steps from each initial input.

\subsubsection{Meandering zonal jet}
The Bickley jet describes a meandering zonal jet flanked above and below by counterrotating vortices \cite{bickley1937lxxiii} and is used to represent an idealized kinematic model of eastward zonal flows such as in Oceanic Gulf Stream or atmospheric polar night jets \cite{rypina2007lagrangian, rypina2007lagrangian}. The model is composed of a steady background flow superimposed with a time dependent perturbation in the form of Rossby waves. The details of the implementation for this experiment are listed in Appendix \href{app:Bickley}{B}. The models were trained on 1200 snapshots and evaluated on 200 snapshots. Model predictions were evaluated over a horizon of 100 future time steps from each initial input.

\subsubsection{Gulf of Mexico Sea Surface Temperature}
Finally, we consider NOAA Optimum Interpolation SST V2 High Resolution Dataset data provided by the NOAA Physical Sciences Laboratory, Boulder, Colorado, USA, from their website at 
https://psl.noaa.govby
\cite{huang2021improvements}. These data are assimilated from various sensing platforms over a long sampling window and interpolated to a regular grid, making it ideal for scientific analysis while reducing sensor bias by referencing multiple sources. We consider the monthly average SST, windowed to the region containing the Gulf of Mexico, which is a common benchmark for similar methods \cite{erichson2019physics}.
The models are trained on 1000 snapshots and evaluated over 500. Model predictions were evaluated over a horizon of 25 future time steps from each initial input, or slightly more than two years.

\subsection{Experimental Overview}
For each experiment we construct both a baseline model ($Baseline$) with loss given by eq. \eqref{eq: AE loss}, and a spectrally-informed model ($SIAE$) with loss given by eq. \eqref{eq: SIAE loss}. We additionally consider two ablated models with partial implementations of the SIAE loss.
These ablation trails consist of a model with only the spectral-informing term ($SIAE_{spectral}$), and a model with only the low-frequency bias term ($SIAE_{\omega}$).

All shared parameters and design choices for these models are identical. In all trials, we use the same model initialization between baseline and spectrally-informed models. In designing the models, we aim to select the simplest suitable architecture, to demonstrate the effectiveness of the spectral-informing mechanism.
All models are constructed using a shallow-encoder/decoder architecture with a single hidden layer and Scaled Exponential Linear Unit (SELU) activation functions; this approach has previously demonstrated good performance on relevant fluid reconstruction tasks \cite{kim2022fast, erichson2020shallow, erichson2019physics}. The input to the model is a single vectorized snapshot of a scalar field (e.g., temperature or viscosity), and the output is a prediction one time step into the future. Longer prediction lengths are generated by autoregressively repeating this process.

\begin{figure}
    \centering
    \includegraphics[width=1.0\columnwidth]{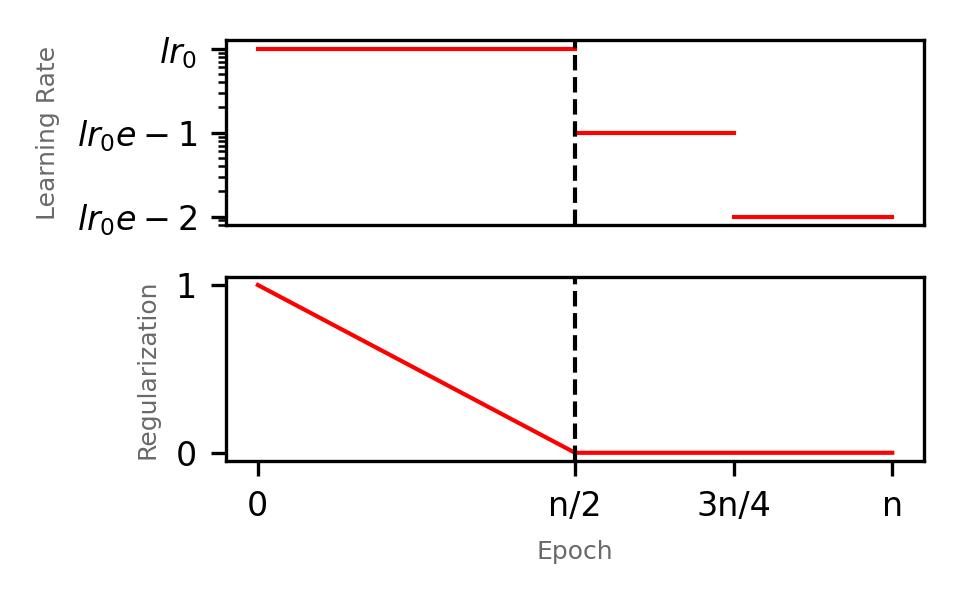}
    \caption{Description of model training procedure, the same learning rate schedule was used for all experiments, where $lr_0$ is the initial learning rate and $n$ is the maximum number of training epochs. The learning rate is decreased by an order of magnitude twice during training. All regularization is applied only during the first half of training, with lower learning rates for fine tuning.}
    \label{fig:training_schematic}
\end{figure}

Hyperparameters were selected via grid search on the training data for learning rate, hidden layer size, and tunable parameters in the regularization functions. The models are trained to convergence, and the same numer of weight updates is used for the spectrally-informed models. For all models the learning rate is decreased twice during training, and regularizations are only applied for the first half of training, as described in Figure \ref{fig:training_schematic}. The regularization factors are scaled linearly from their initial value to zero at half of the full training epochs. This allows the models to learn with the described biases but fine tune without them. Model parameters are described in Appendix \href{app:parameters}{A}.

\subsection{Metrics and Summary of Results}

We consider five metrics, which together describe both the performance of the model and the impact of the physical intuitions applied to the spectrally-informed model. For all evaluations, predictions are generated on a separate test set of data over a number of trajectories, the size of the test dataset and length of prediction trajectories for each case are provided in the data overview. The five metrics are:
\begin{enumerate}
    \item \textbf{Time average mean squared prediction error ($MSE$)}: computed as average ratio of prediction errors between the baseline model and spectrally-informed model, over each prediction trajectory in the test set.
    We normalize the MSE of all models to the baseline MSE for easier comparison across experiments, a value below 1.0 indicates improved average prediction accuracy over the trajectories compared to the baseline model.
    \item \textbf{Structural Similarity Index Measure ($SSIM$)}: is a metric describing perceived similarity between images via localized FFT comparisons \cite{wang2004image}. We compute the SSIM for the baseline and SIAE models over all prediction steps and average; a higher value indicates greater perceived similarity, up to a maximum of one.
    \item \textbf{Average eigenfrequency ($\langle\omega_j\rangle$)}: the goal of the frequency regularization term ($\pazocal{L}_\omega$) is to prioritize learning models with dynamics which operate along lower frequencies, which are assumed to be more energy containing and generalizable. To evaluate this, we record the average eigenfrequency for the linear dynamics of each fully trained model. These values are reported normalized to the Nyquist frequency, where a value of 1 indicates the highest discernible frequency at half the sampling rate, and 0 indicates a static term. The dynamics are randomly initialized using a Xavier uniform distribution, which on average results in a mean eigenfrequency of 0.5 at initialization \cite{o2016eigenvectors}.
    \item \textbf{Maximum growth/decay ($\gamma_{max}$)}: we record the maximum magnitude of the linear operator's eigenvalues to characterize the stability of the learned dynamics.
    All models have a stability regularization term developed in \cite{erichson2019physics}. A negative value indicates decaying or stable dynamics and a positive value indicates exponential growth. For realistic flows, model utility is expected to degrade over long prediction lengths relative to the correlations in the flow, limiting the relevant time horizon of predictive models. Therefore, although small positive values do indicate unstable dynamics, on practical time scales these may be sufficient.
    \item \textbf{Spectral error ($E_{spectrum}$}): we evaluate the spectral loss of the final model as descried in eq. \eqref{eq: spectral loss} and normalize to the baseline for comparison. Unlike the regularization used in training, this is computed over the entire test set. A value below 1.0 indicates an improved match to the prior spectrum over the baseline mode. The prior spectrum is determined from the data in all cases.
\end{enumerate}

\begin{table}[htbp]
  \centering
  \caption{Summary of Results}
  \label{tab:results}
  \begin{tabular}{lccccc}
    \toprule
    \textbf{Experiment} & \multicolumn{5}{c}{\textbf{Metrics}} \\
    \cmidrule{2-6}
    & \textbf{$MSE$} & $SSIM$ & \textbf{$\langle\omega_j\rangle$} & $\gamma_{max}$ &  
    \textbf{$E_{spectrum}$} \\
    \midrule
    \multicolumn{5}{l}{\textbf{Periodic vortex shedding}} \\
    Baseline & 1.00 & 0.734 & 0.475 & 0.051 & 1.00 \\
    $SIAE_{spectral}$ & 0.489 & 0.930 & 0.503 & 0.147  & 0.020 \\
    $SIAE_{\omega}$ & 0.934 & 0.756 & \textbf{0.390} & -0.001 & 0.029 \\
    $SIAE$ & \textbf{0.447} & \textbf{0.937} & 0.484 & 0.065 & \textbf{0.014} \\
    \midrule
    \multicolumn{5}{l}{\textbf{Meandering zonal jet}} \\
    Baseline & 1.00 & 0.503 & 0.508 & -0.147  & 1.00 \\
    $SIAE_{spectral}$ & 0.871 & 0.535 & 0.504 & -0.103 & 0.843 \\
    $SIAE_{\omega}$ & 0.784 & 0.575 & \textbf{0.130} & 0.001 & \textbf{0.235} \\
    $SIAE$ & \textbf{0.661} & \textbf{0.642} & 0.163 & 0.007 & 0.412 \\
    \midrule
    \multicolumn{5}{l}{\textbf{Sea surface temperature}} \\
    Baseline & 1.00 & 0.611 & 0.478 & -0.158 & 1.00 \\
    $SIAE_{spectral}$ & 0.783 & 0.703 & 0.453 & -0.005  & 0.741 \\
    $SIAE_{\omega}$ & 0.765 & 0.706 & \textbf{0.110} & 0.009 & 0.900 \\
    $SIAE$ & \textbf{0.731} & \textbf{0.718} & 0.224 & -0.001 & \textbf{0.416} \\
    \bottomrule
  \end{tabular}
\end{table}

\begin{figure}
    \centering
    \includegraphics[width=1.0\columnwidth]{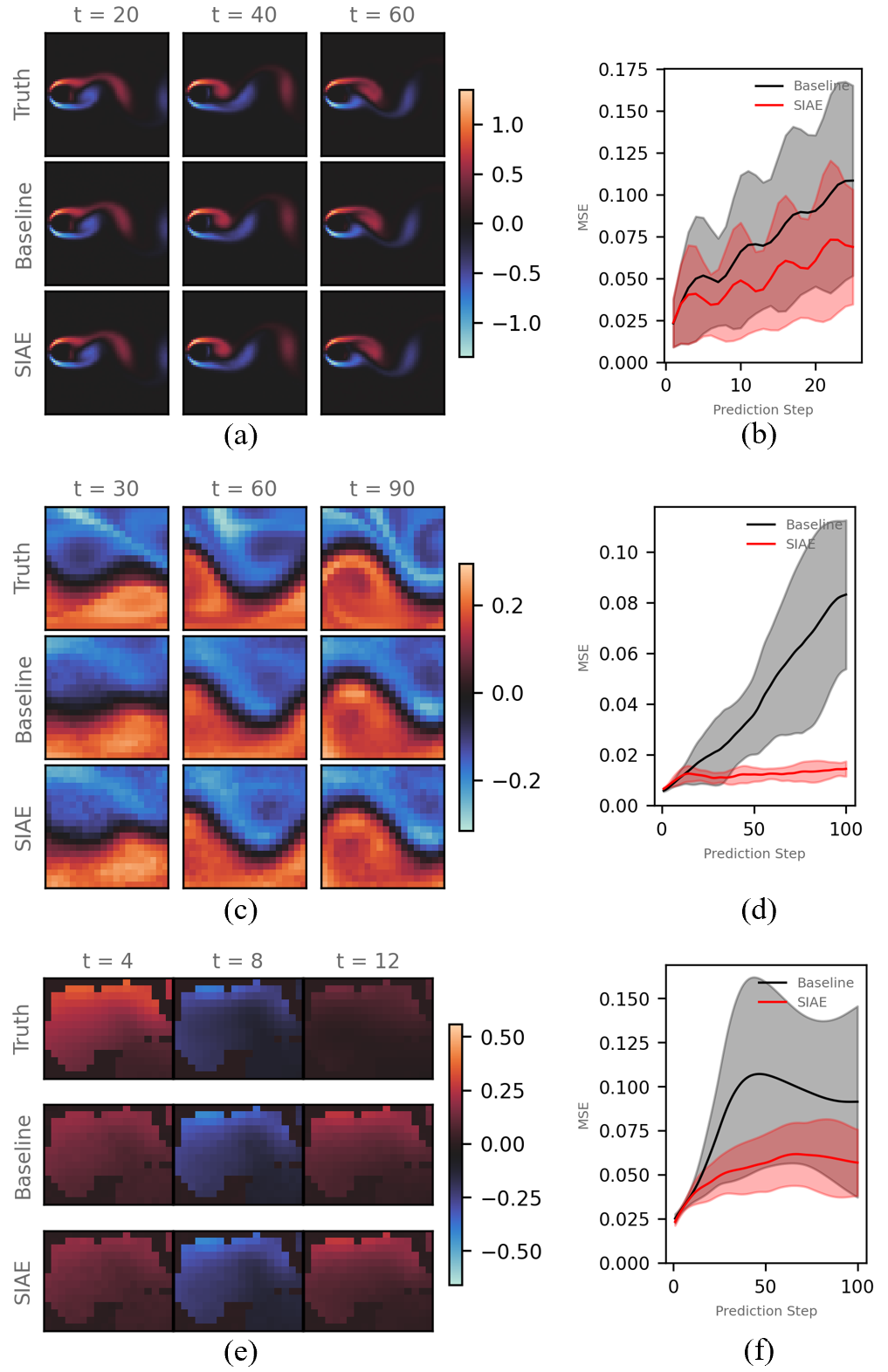}
    \caption{Overview of experimental results, showing sample predictions and MSE over the test prediction horizon for: flow past a cylinder, (a), (b), Bickley jet, (c), (d), and Gulf of Mexico SST, (e), (f). All MSE plots show $2\sigma$ confidence intervals.}
    \label{fig:results overview}
\end{figure}

These metrics are summarized for each experiment in table \ref{tab:results}; visual demonstrations for the SIAE and baseline models in each experiment are shown in Figure \ref{fig:results overview}.

Overall, we find that SIAE model and both ablated implementations result in improved performance over the baseline model in terms of MSE and SSIM for all three experiments. In all cases, the full implementation ($SIAE$) with both regularization terms resulted in the best predictive performance, suggesting a complementary effect of low-frequency and spectral matching priors. The detailed results reveal that these performance improvements are a result of improved long-term prediction accuracy; the regularized models offer no significant improvement over the baseline model in short-term predictions in all cases. This is a somewhat intuitive result that highlights the potential tradeoffs between physically-informed priors and purely data-driven learning. The baseline model is trained on short term predictions and results in highly accurate models within this window; by incorporating physical knowledge, the learned dynamics better match the underlying structure and result in improved long term accuracy.

We also confirm the successful application of the regularization techniques. Nearly all models with low-frequency bias included ($SIAE$ and $SIAE_\omega$) resulted in dynamics with lower average eigenfrequency than the baseline model, with the one exception being the SIAE model for flow past a cylinder. As demonstrated in Figure \ref{fig:low f bias demo}, the baseline models result in no substantial changes to the mean eigenfrequency from the initialization of $\langle\omega_j\rangle = 0.5$. The introduction of the low-frequency bias resulted in a reduction as high as 75.0\% for modeling SST.
We note that lower eigenfrequencies in the learned dynamics does not inherently indicate a more model which better matches the underlying physics, but may offer particular advantages by corresponding to higher energies in the power spectrum which are often more general and informative and accordingly more amenable to sensing and control.
We note this result has potential in a much broader class of dissipative processes which result in a decaying power spectrum.

We also observe a decrease in spectral error across all models with a spectral loss term ($SIAE$ and $SIAE_{spectral}$), and interestingly, some models with only the low-frequency bias ($SIAE_\omega$). This both demonstrates the ability to inform the model to closer match the prior spectrum, and indicates that this effect results in improved performance if the prior is well selected. For both the Bickley jet and SST cases, the incorporation of the low-frequency bias further reduces the spectral error. The low-frequency term alone resulted in reduced spectral error, likely due to shifting high power, high frequency eigenvalues to lower frequencies which have higher power in the prior spectrum, resulting in a better match and reduced spectral error.

\begin{figure}
    \centering
    \includegraphics[width=1.0\columnwidth]{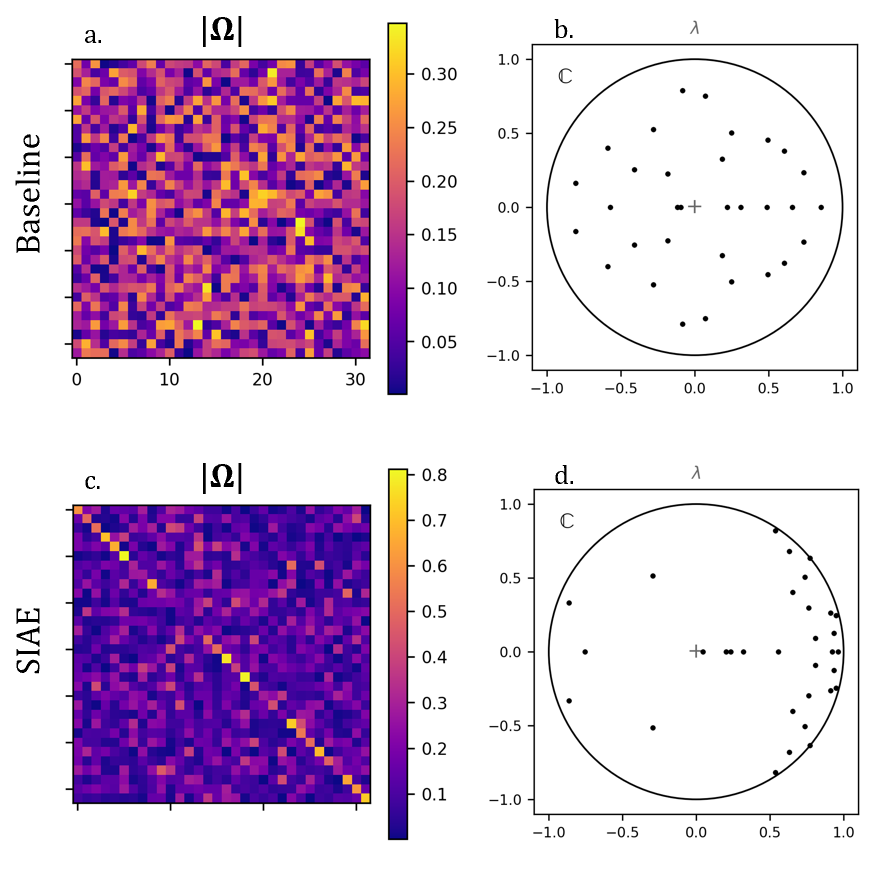}
    \caption{Comparison of linear dynamics of fully trained models modeling SST data for (a) the baseline model and (c) the SIAE model and the corresponding eigenvalues for (b) the baseline model and (d) the SIAE model. We observe that the SIAE model learns more persistent features with lower frequency as indicated by a clear diagonal in the dynamics and lower average phase and higher magnitude of eigenvalues.}
    \label{fig:sst_eigs}
\end{figure}

We note that in all cases the baseline model is stable or less unstable than the SIAE model, as indicated by the maximum eigenvalue growth, $\gamma_{max}$. This indicates that while stability-promoting priors are theoretically supported and provide useful properties, they will not necessarily result in improved performance, as the constraint is often satisfied implicitly.

Figure \ref{fig:sst_eigs} demonstrates the learned linear dynamics and corresponding eigenvalues for the baseline and SIAE models trained on SST data. We observe greater structure in the dynamics, particularly in the form of an apparent diagonal which corresponds to self-similarity over time. Examining the eigenvalues directly reveals the learned dynamics operate on lower frequency (corresponding to phase) and are more persistent (magnitude closer to 1). While the frequency is directly informed via the low-frequency bias, the improved persistence is emergent from the training process, and may suggest a better representation of underlying flow features.

In the era of modern machine learning, larger models trained on more data generally provide the best performance as measured by error over the test data. Although the spectral-informing mechanism resulted in improved performance over the baseline model with the same configuration, we do not claim state-of-the-art performance on the modeling tasks considered. Rather, we have demonstrated a proof-of-concept for an approach to incorporate physical knowledge on the spectral properties of a system as interpretable biases in the learning process. This method may be particularly useful in cases on limited data or computation, as is often the case in real-time applications.

\section{Conclusions}

In this work, we have developed and demonstrated a mechanism for incorporating prior spectral knowledge into the learning process of a KAE model. By modifying the training process, we demonstrate the ability to learn dynamics which evolve on lower frequencies, are more energy containing and generalizable, and better match a prior known energy spectrum. The resulting SIAE approach is shown to be an effective improvement over a baseline model without these physical priors on prototypical fluid flow examples. While several methods have been proposed for incorporating physical knowledge into KAE models, to our knowledge this is the first approach which explicitly utilizes domain specific frequency spectrum characteristics. Our results indicate the potential of incorporating spectral priors into data-driven models, even in the form of the very general low-frequency bias.

The resulting SIAE approach is simple to implement and general. The limitations of this method arise from the assumptions underlying its development. These may not be met in systems without underlying low-rank structure, with a continuous spectrum, with important features at high frequencies, or when spectral priors are not known. While we present an effective method for incorporating implicit spectral priors as optimization goals, future work should further investigate the role of initialization, and consider an explicit form of these constraints, i.e., by clamping eigenfrequencies at a maximum value. Future work should investigate the potential of the SIAE method for control applications, which are inherently limited by bandwidth, and could benefit from models which evolve along lower frequencies. Overall,  efficient fluid flow models are vital to real time applications in fluid environments, and we have demonstrated that these models can be improved through the incorporation of prevalent spectral insights from the physical systems of interest.

\begin{acknowledgments}
BDS thanks Austin McDaniel, Christopher Wilcox, Matthew Kemnetz of AFRL Directed Energy Directorate, Neel Dhulipala from the Olin College of Engineering, and Dhilan Desai from the University of Illinois Urbana-Champaign for helpful discussions. \par Approved for public release; distribution is unlimited. Public Affairs release approval \#AFRL-2024-20242195. The views expressed are those of the authors and do not necessarily reflect the official policy or position of the Department of the Air Force, the Department of Defense, or the U.S. government. 
\end{acknowledgments}

\section*{Data Availability Statement}

All data used in this report are publicly available. The DNS flow past a cylinder data were retrieved from CGL at ETH Zurich, the monthly SST data were retrieved from NOAA Boulder, and the Bickley jet experiment was created following the description in Appendix \href{app:Bickley}{B} and can be provided by the authors upon reasonable request.

\section*{References}
\bibliography{bib}

\appendix
\section*{Appendix A. Model Parameters}
\label{app:parameters}
The model parameters for all trials are listed in \ref{tab:hyperparameters}. The tunable SIAE weighting parameters were selected based on simple grid search over the training data.

\begin{table}[htbp]
    \centering
    \caption{Neural Network Hyperparameters}
    \label{tab:hyperparameters}
    \begin{tabular}{lccc}
        \toprule
        \multirow{2}{*}{\textbf{Parameter}} & \multicolumn{3}{c}{\textbf{Experiment}} \\ 
        & Vortex Shedding & Bickley Jet & Gulf SST \\
        \midrule
        Epochs & 100 & 50 & 100 \\
        $lr_0$ & 5e-4 & 5e-3 & 5e-3 \\
        $\alpha_{spectral}$ & 1.0 & 1e-4* & 1e-1 \\
        $\alpha_\omega$ & 1e-3 & 1e-2 & 1e-2 \\
        $\alpha_\gamma$ & 1e-1 & 1e-1& 1e-2 \\
        Hidden Layer Size & 64 & 128 & 256 \\
        Latent Dimension ($h$) & 32 & 64 & 32 \\
        Prediction Horizon & 100 & 100 & 25 \\
        \bottomrule
        \multicolumn{4}{c}{* a value of 1e-10 was used in the spectral bias only trial}
    \end{tabular}
\end{table}

\section*{Appendix B. Bickley Jet Formulation}
\label{app:Bickley}
The total stream function is modeled as
\begin{equation}
    \psi(x,y,t) = \psi_0(y)+\psi_1(x,y,t)
\end{equation}
with steady background flow
\begin{equation}
    \psi_0(y)=-ULtanh(\frac{y}{L})
\end{equation}
and traveling Rossby wave perturbation
\begin{equation}
    \psi_1(x,y,t) = UL \sech^2(\frac{y}{L})\Re[\sum^3_{n=1}(f_n(t)e^{ik_nx})]
\end{equation}
where $f_n(t) = \epsilon_n \exp(-ik_nc_nt)$. The velocity field of the Bickley jet is computed as $u=-\frac{\partial\psi}{dy}$ and $v=\frac{\partial\psi}{dx}$. Our implementation exactly follows \cite{hadjighasem2017critical, salam2022online} with 
scaled parameters $U = 5.4138$, $L = 1.77$, $c1 = 0.1446U$, $c2 = 0.2053U$, $c3 =0.4561U$, $\epsilon_1 = 0.075$, $\epsilon_2= 0.4$, $\epsilon_3 = 0.3$, $r = 6.371$, $k1 = 2/r$,$k2 = 4/r$,$k3 = 6r$. We sample 9900 points uniformly on a grid $[0,20] \times [-3,3]$. To generate the data we numerically advect a scalar field with a periodic $x$ boundary. At each time step the first column of the scalar field is set to $s(0,y) = y$, giving a vertical gradient. The advection is applied $t \in [0,40]$ with step size 0.1, and the final data is taken by windowing out the temporal and spatial transience caused by the initial condition. The training data is downsampled to 24 by 24.

\section*{Appendix C. Sensitivity of Tuning Parameters}
\label{app:Sensitivity}
The spectral-informing implementation introduces three new tunable parameters to the autoencoder loss that balance the relative training goals. It is desirable that these parameters be robust, i.e. a small change in parameter value results in a small change in model performance. Since these parameters would likely be tuned on training data in application, it is also preferred that similar performance can be achieved within an order of magnitude variation of the optimal value.
Previous work has shown the stability loss parameter is highly robust \cite{erichson2019physics}.

\begin{figure}
    \centering
    \includegraphics[width=1.0\columnwidth]{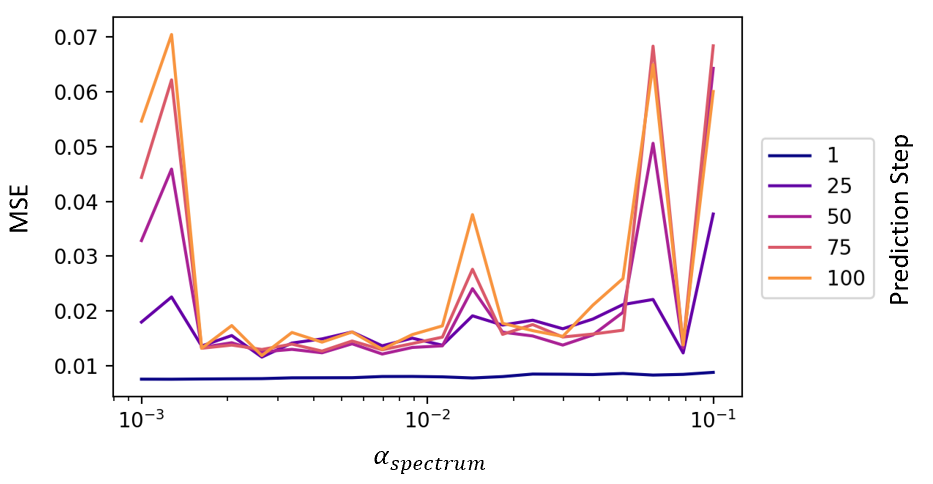}
    \caption{The sensitivity of model performance to spectral regularization shows generally robust performance within $\pm1$ order of magnitude from grid-search selected value. These results show average prediction errors for flow past a cylinder at time horizons of 1, 25, 50, 75, and 100 time steps into the future.}
    \label{fig:sensitivity alpha spectrum}
\end{figure}

Figure \ref{fig:sensitivity alpha spectrum} demonstrates the sensitivity of model prediction accuracy at different time lengths to the $\alpha_{spectrum}$ parameter for the vortex shedding dataset. We compare models trained with 20 different values of $\alpha_{spectrum}$ within $\pm1$ order of magnitude of the selected value of $10^{-2}$ resulting from a grid search on the training data. We see that within this range there is a region of good performance with one exception, and that the performance diverges at significantly higher or lower values. The robustness of this parameter likely depends on the details of training, architecture, and problem, however for the all cases in this work we see similar results, suggesting a reasonably wide range of values result in similar, suitable performance.

\section*{Appendix D. Latent Space Power Spectra}
\label{app:Latent Spectra}

\begin{figure}
    \centering
    \includegraphics[width=1.0\columnwidth]{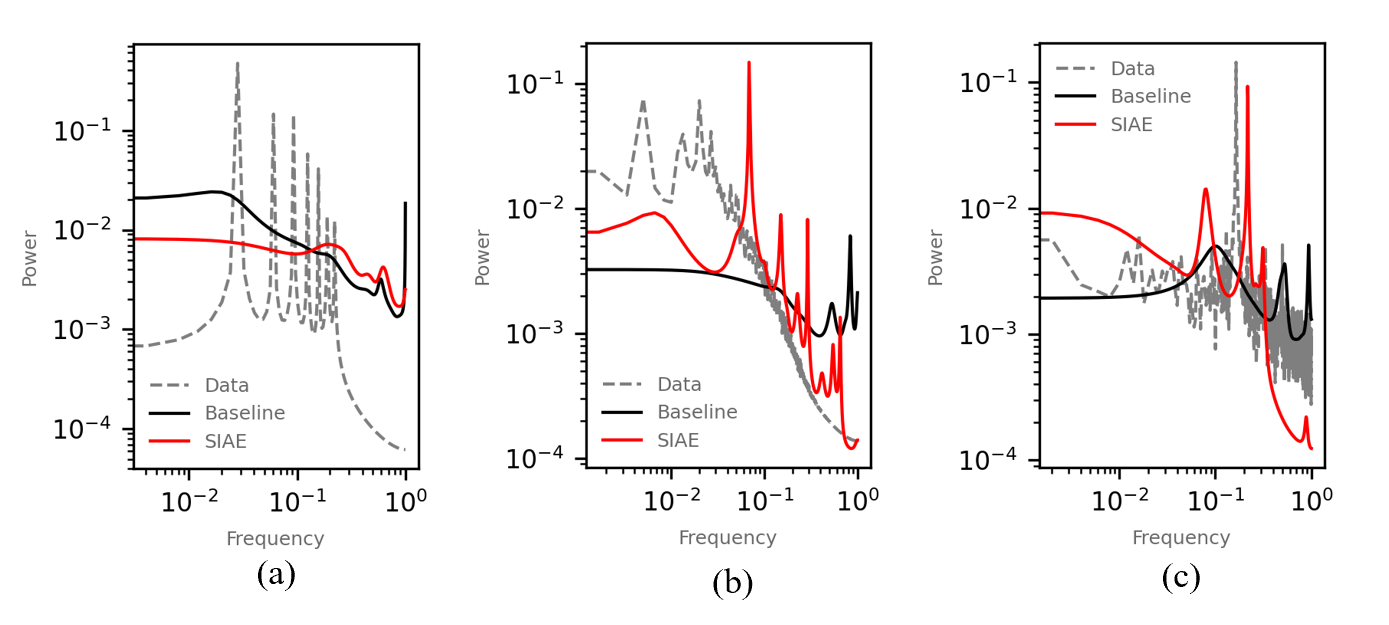}
    \caption{Power spectra for the SIAE model (red) and baseline (black), compared to temporal power spectrum computed from data (grey) in each experiment computed over trajectories in the latent space for: (a) flow past a cylinder, (b) Bickley jet, (c) SST.}
    \label{fig:latent PSDs}
\end{figure}

To demonstrate the spectrum of the learned dynamics, we approximate $\hat{E}_{\bm{\Omega}}$ by computing the FFT over a trajectory of states in the latent space from a given input. For each sample we apply the normalized linear dynamics $\Bar{\bm{\Omega}}^t$ with $t = \{1, 2, \cdots, p\}$ to give a trajectory in the latent space for that snapshot $\mathbf{Y}_k \in \mathbb{R}^{p\times h}$,
\begin{equation}
    \mathbf{Y}_k = [\Bar{\bm{\Omega}}^1\mathbf{y}_k, \Bar{\bm{\Omega}}^2\mathbf{y}_k, \cdots, \Bar{\bm{\Omega}}^p\mathbf{y}_k,]
\end{equation}
where $\Bar{\bm{\Omega}} = \bm{\Omega} / max_i |\lambda_i|$ to ensure stability over the prediction interval.
This process is repeated for each $k = \{1, 2, \cdots, T\}$, where $T$ is the length of the training data.
For each dimension in the latent space, we compute the real-valued FFT, and estimate the power spectrum of the linear operator by spatially averaging these spectra, 
\begin{equation}
    \Tilde{E}_{\bm{\Omega}} = \langle\lvert\hat{\mathbf{Y}}_k\rvert\rangle
\end{equation}
where $\hat{\cdot}$ indicates a temporal Fast Fourier Transform and $\langle\cdot\rangle$ indicates a spatial average and unit energy normalization. This method for computing the latent space power spectrum captures both the relative power and the relative growth/decay over the prediction trajectory, and is therefore useful for analysis of the resulting model, but is too computationally expensive for practical implementation during training.

The resulting spectra are shown for each experiment in Figure \ref{fig:latent PSDs}. We observe that in all cases the baseline model results in a relatively flat power spectrum. In both the Bickley jet and SST cases, the SIAE model results in higher relative power at lower frequencies, and appears to better match the empirical spectrum computed from the data snapshots. This is not the case in the SIAE model for flow past a cylinder, where the low-frequency bias was not effective in the full SIAE model, as recorded in table \ref{tab:results}.

\end{document}